\documentclass[jkps,preprint,fleqn,showpacs,showkeys]{revtex4}
\usepackage{graphicx}
\usepackage{amssymb}
\usepackage{amsmath}
\usepackage{bm}
\begin{document}
\setcounter{page}{0}
\title[]{Teleparallel Energy-Momentum Distribution of Locally Rotationally Symmetric  Spacetimes}
\author{M. Jamil \surname{Amir}}
\email{mjamil.dgk@gmail.com}
\author{Tahir \surname{ Nazir}}
\email{tahirnazir66@yahoo.com}
\affiliation{Department of Mathematics, University of Sargodha,\\
Sargodha-40100, Pakistan}

\date[]{Received }

\begin{abstract}
In this paper, we explore the energy-momentum distribution of
locally rotationally symmetric (LRS) spacetimes in the context of the
teleparallel theory of gravity by considering the three metrics, I, II and III,
representing the whole class of LRS sapcetimes. In this regard, we use the
teleparallel versions of the Einstein, Landau-Lifshitz,
Bergmann-Thomson, and M$\ddot{o}$ller prescriptions. The results
show that the momentum density components for the Einstein,
Bergmann-Thomson, and M$\ddot{o}$ller prescriptions turn out to be same
in all cases of the metrics I, II and III, but are different from those of the Landau-
Lifshitz prescription, while the energy components remain the same for these
three prescriptions only in all possible cases of the metrics I and II.
We mention here that the M$\ddot{o}$ller energy-momentum distribution
is independent of the coupling constant $\lambda$; that is, these results are
valid for any teleparallel models.
\end{abstract}



\maketitle

\section{INTRODUCTION}

\quad One of the most interesting, but challenging, problems in Einstein's theory of
general relativity (GR) is the localization of energy. This problem still
needs a definite answer due to its unusual nature and the various viewpoints on it.
Among all available theories of gravitation in the literature, GR has been
accepted as the true theory of gravitation as many physical aspects of nature
have been experimentally verified in the context of this theory. However,
the localization of energy and momentum [1] is still an open,
unresolved and disputed problem in GR. In GR, many attempts have been made to
resolve this problem, but no definition has generally been accepted till now.

As a pioneer, Einstein  used the notion of an energy-momentum    
complex to solve this problem [2]. Following Einstein, many scientists
like Landau-Lifshitz [3], Papapetrou [4], Bergmann-Thomson [5],
Tolman [6], Weinberg [7] and M$\ddot{o}$ller [8] have introduced
their own energy-momentum complexes. All these prescriptions,
except M$\ddot{o}$ller's, are restricted to carrying out the
calculations in Cartesian coordinates only for the sake of
physical results. Also, we cannot define angular momentum with
the help of all these prescriptions. Misner et al. [1] showed that
energy can only be localized in spherical systems, but very soon after that
Cooperstock and Sarracino [9] proved that if energy is localizable
for spherical systems, then it can be localized in any coordinate
system. Bondi [10] argued that a non-localizable form of energy is
not admissible in GR. After this, the idea of quasi-local energy was introduced by Penrose and
other scientists [11-13]. In this method, one can use any coordinate system
while finding the quasi-local masses to obtain the energy-momentum of a
curved spacetime. Bergqvist [14] considered seven different definitions of
quasi-local masses and showed that no two of these definitions gave the same
results. Chang et al. [15] proved that every energy-momentum complex could be
associated with a particular Hamiltonian boundary term. Thus, the
energy-momentum complexes may also be considered as quasi-local. Xulu
[16-17] extended this investigation and found the same energy distribution
in the cases of the Melvin magnetic and the Bianchi type I universes.

Virbhadra and his collaborators [18-21] verified for
asymptotically-flat spacetimes that different energy-momentum
complexes could give the same result for a given spacetime. They
also found encouraging results for the case of asymptotically
non-flat spacetimes by using different energy-momentum complexes.
Aguirregabiria et al. [22], by using the Einstein, Landau
Lifshitz, Papapetrou, Bergmann, and Weinberg (E,LL,P,B,W)
prescriptions, showed that the energy distributions within a
Kerr-Schild metric were the same. Virbhadra [23] found that these five
different prescriptions (E,LL,P,B,W) did not give the same results for
the most general non-static spherically-symmetric spacetime. One
of the authors found several examples that did not provide
the same result for different prescriptions [24] . The results found in
[17, 19, 21 and 23-25] lead us to know that the energy distribution in
M$\ddot{o}$ller's prescription is different from Einstein's energy
for some particular spacetimes, including the Schwarzschild spacetime.

Some authors [26-30] argued that this problem of energy might be settled in
the context of the teleparallel theory (TPT) of gravity. They showed that
energy-momentum could also be localized in the framework of this theory. The results of the two theories have been shown to agree with each other.
Vargas [28] found that the total energy of the closed
Friedmann-Robertson-Walker spacetime was zero by using the teleparallel versions
of the Einstein and the Landau-Lifshitz complexes. This agrees with the result
obtained by Rosen [31] in GR. Salti and his co-workers [29] considered some
particular spacetimes and calculated the energy-momentum densities by using
different prescriptions both in GR and TPT and they found the similar results.

Sharif and Amir [32-38] evaluated the energy-momentum
distribution of the Lewis-Papapetrou spacetime by using the TP version of
M$\ddot{o}$ller's prescription and found that the results did not
agree with those available in the context of GR. We use the
TP versions of the M$\ddot{o}$ller, Einstein,
Landau-Lifshitz and Bergmann-Thomson prescriptions to find the
energy-momentum distribution of this metric and compare the
results with those already found in GR. However, the
energy-momentum density components become the same in both
theories under certain assumptions. They also discussed the
energy-momentum of  static, axially-symmetric spacetimes in the
framework of teleparallel gravity (TPG). For this purpose, they used the TP versions of
the Einstein, Landau-Lifshitz, Bergmann-Thomson and M$\ddot{o}$ller
prescriptions. A comparison of the results shows that the energy
densities are different, but the momentum turns out to be constant in
each prescription. This is exactly similar to the results
available in literature when using the framework of GR.

Sharif and his collaborators [39-41] found that the
results for the energy exactly coincided with several prescriptions in GR.
Interestingly, our results exactly coincide with different energy-momentum
prescriptions in GR. The constant momentum shows consistency with
the results available in GR and TPG. Recently, Amir et al. [42] explored the energy-momentum
distribution of non-static plane symmetric spacetimes in GR and (TPG).

The scheme of this paper is as follows. In Section \textbf{II}, we
give some basics of the TPT and the TP versions of the Einstein,
Landau-Lifshitz, Bergmann-Thomson and M$\ddot{o}$ller
prescriptions. Section \textbf{III} is devoted to evaluating of
the energy-momentum density components for locally rotationally-
symmetric spacetimes. In the last section, we shall summarize the
results.

\section{TELEPARALLEL GRAVITY AND ENERGY-MOMENTUM COMPLEXES}
\quad First, we briefly outline the main points of the TPT.
The basic entity of the TPG is
the non-trivial tetrad  ${h^{a}}_{\mu }$, whose inverse is
denoted by ${h_{a}}^{\nu }$, satisfying the relations [34]
\begin{equation}
{h^{a}}_{\mu }{h_{a}}^{\nu }={\delta _{\mu }}^{\nu };\quad
\texttt{and} \quad \ {h^{a}}_{\mu }{h_{b}}^{\mu }={\delta
^{a}}_{b}.
\end{equation}
The theory of TPG is described by the Weitzenb$\ddot{o}$ck connection, which is given
as
\begin{equation}
{\Gamma ^{\theta }}_{\mu \nu }={{h_{a}}^{\theta }}\partial _{\nu
}{h^{a}}_{\mu }
\end{equation}
and is obtained due to the condition of absolute parallelism [35]. This
implies that the spacetime structure underlying a translational gauge theory
is naturally endowed with a teleparallel structure [35-36]. In this paper,
the Latin alphabet $(a,b,c,...=0,1,2,3)$ will be used to denote the tangent
space indices and the Greek alphabet $(\mu ,\nu ,\rho ,...=0,1,2,3)$ to
denote the spacetime indices. The Riemannian metric in TPT arises as a byproduct [35] of the tetrad field given by
\begin{equation}
g_{\mu \nu }=\eta _{ab}{h^{a}}_{\mu }{h^{b}}_{\nu },
\end{equation}%
where $\eta _{ab}$ is the Minkowski spacetime such that $\eta
_{ab}=diag(+1,-1,-1,-1)$.

 In TPT, the gravitation is attributed to torsion
[36], which plays the role of a force here. For the Weitzenb$\ddot{o}$ck
spacetime, the torsion is defined as [37]
\begin{equation}
{T^{\theta }}_{\mu \nu }={\Gamma ^{\theta }}_{\nu \mu }-{\Gamma ^{\theta }}%
_{\mu \nu },
\end{equation}%
which is antisymmetric in nature. Due to the requirement of absolute
parallelism, the curvature of the Weitzenb$\ddot{o}$ck connection vanishes
identically [34]. The Weitzenb$\ddot{o}$ck connection and the Christoffel
symbols satisfy the following relation:
\begin{equation}
{{\Gamma ^{0}}^{\theta }}_{\mu \nu }={\Gamma ^{\theta }}_{\mu \nu }-{%
K^{\theta }}_{\mu \nu },
\end{equation}%
where ${{\Gamma ^{0}}^{\theta }}_{\mu \nu }$ are the Christoffel symbols and
${K^{\theta }}_{\mu \nu }$ denotes the \textbf{contorsion tensor} and is
given by
\begin{equation}
{K^{\theta }}_{\mu \nu }=\frac{1}{2}[{{T_{\mu }}^{\theta }}_{\nu }+{{T_{\nu }%
}^{\theta }}_{\mu }-{T^{\theta }}_{\mu \nu }].
\end{equation}%

   The teleparallel versions of the Einstein, Landau-Lifshitz and Bergmann
energy-momentum complexes, by setting $c=1=G$, are, respectively, given by
[28]
\begin{eqnarray}
hE_{\nu }^{\mu } &=&\frac{1}{4\pi }\partial _{\lambda }({U_{\nu }}^{\mu
\lambda }),  \nonumber \\
hL^{\mu \nu } &=&\frac{1}{4\pi }\partial _{\lambda }(hg^{\mu \beta }{%
U_{\beta }}^{\nu \lambda }),  \nonumber \\
hB^{\mu \nu } &=&\frac{1}{4\pi }\partial _{\lambda }({g^{\mu \beta }U_{\beta
}}^{\nu \lambda }),
\end{eqnarray}%
where ${U_{\nu }}^{\mu \lambda }$ is  Freud's superpotential given as
\begin{equation}
{U_{\nu }}^{\mu \lambda }=h{S_{\nu }}^{\mu \lambda }.
\end{equation}%
Here, $S^{\nu \mu \lambda }$ is a tensor quantity that is skew symmetric in
its last two indices and is defined as
\begin{equation}
S^{\nu \mu \lambda }=m_{1}T^{\nu \mu \lambda }+\frac{m_{2}}{2}(T^{\mu \nu
\lambda }-T^{\lambda \nu \mu })+\frac{m_{3}}{2}(g^{\nu \lambda }{T^{\beta
\mu }}_{\beta }-g^{\mu \nu }{T^{\beta \lambda }}_{\beta }),
\end{equation}%
where $m_{1},~m_{2}$ and $m_{3}$ are three dimensionless coupling constants
of TPG [35]. we mentioned here that $hE_{0}^{0}$, $hL^{00}$ and $hB^{00}$ are
the energy densities, $hE_{i}^{0},~hL^{0i}$ and $hB^{0i}~(i=1,2,3)$ are the
momentum densities, and $hE_{0}^{i},~hL^{i0}$ and $hB^{i0}$ are the energy current
densities of the Einstein, Landau-Lifshitz and Bergmann prescriptions,
respectively.

The Teleparallel equivalent of GR may be obtained by considering
the following particular choice [35]:
\begin{equation}
m_{1}=\frac{1}{4},\quad m_{2}=\frac{1}{2},\quad m_{3}=-1.
\end{equation}%
The superpotential of  M$\ddot{o}$ller's tetrad theory was given by Mikhail
et al. [26] as
\begin{equation}
{U_{\mu }}^{\nu \beta }=\frac{\sqrt{-g}}{2\kappa }P_{\chi \rho \sigma
}^{\tau \nu \beta }[V{^{\rho }}g^{\sigma \chi }g_{\mu \tau }-\lambda g_{\tau
\mu }K^{\chi \rho \sigma }-(1-2\lambda )g_{\tau \mu }K^{\sigma \rho \chi }],
\end{equation}%
where
\begin{equation}
P_{\chi \rho \sigma }^{\tau \nu \beta }={\delta _{\chi }}^{\tau }g_{\rho
\sigma }^{\nu \beta }+{\delta _{\rho }}^{\tau }g_{\sigma \chi }^{\nu \beta }-%
{\delta _{\sigma }}^{\tau }g_{\chi \rho }^{\nu \beta },
\end{equation}%
with $g_{\rho \sigma }^{\nu \beta }$ being a tensor quantity and being defined as
\begin{equation}
g_{\rho \sigma }^{\nu \beta }={\delta _{\rho }}^{\nu }{\delta _{\sigma }}%
^{\beta }-{\delta _{\sigma }}^{\nu }{\delta _{\rho }}^{\beta },
\end{equation}%
$K^{\sigma \rho \chi }$ is the contortion tensor as given by Eq. (6), $g$ is the
determinant of the metric tensor $g_{\mu \nu }$, $\lambda $ is the free
dimensionless coupling constant of TPG, $\kappa $ is the Einstein constant
and $V_{\mu }$ is the basic vector field given by
\begin{equation}
V_{\mu }={T^{\nu }}_{\nu \mu }.
\end{equation}%
Now, we can write the M$\ddot{o}$ller energy, momentum, and energy current
densities as follows:
\begin{equation}
\Xi _{\mu }^{\nu }=U_{\mu }^{\nu \rho },_{\rho },
\end{equation}%
where the comma means ordinary differentiation. Further, $\Xi _{0}^{0},~\Xi _{i}^{0}$
and $\Xi _{0}^{i}$ are the energy, momentum, and energy current densities,
respectively, in M$\ddot{o}$ller's prescription.

\section{EERGY-MOMENTUM DISTRIBUTION OF LRS SPACETIMES}

\quad Many authors [27-30] studied extensively the LRS spacetimes that contain
well-known exact solutions of the Einstein's field equations. They admit a group of motions $G_{4}$
acting multiply transitively on three dimensional non-null orbits, spacelike (%
$S_{3}$) or timelike ($T_{3}$), and the isotropy group is a spatial rotation.
These spacetimes are represented by three families of metrics given as
[26-27]
\begin{eqnarray}
ds^{2} &=&\epsilon \lbrack
-dt^{2}+A^{2}(t)dx^{2}]-B^{2}(t)dy^{2}-B^{2}(t)\Sigma ^{2}(y,k)dz^{2}, \\
ds^{2} &=&\epsilon \lbrack
-dt^{2}+A^{2}(t)dx^{2}]-e^{2x}B^{2}(t)(dy^{2}+dz^{2}),\\
ds^{2} &=&\epsilon \lbrack -dt^{2}+A^{2}(t)\{dx-\Lambda
(y,k)dz\}^{2}]-B^{2}(t)dy^{2}  \nonumber \\
&&-B^{2}(t)\Sigma ^{2}(y,k)dz^{2},
\end{eqnarray}%
where $k=-1,0,1,~\epsilon =\pm 1$,
\[
\Sigma =\left\{
\begin{array}{l}
\sin y,\quad ~~k=1, \\
y,\quad \quad \quad k=0, \\
\sinh y,\quad k=-1, \\
\end{array}%
\right.
\]%
and
\[
\Lambda =\left\{
\begin{array}{l}
\cos y,\quad ~~k=1, \\
\frac{y^{2}}{2},\quad \quad ~~k=0, \\
\cosh y,\quad k=-1. \\
\end{array}%
\right.
\]%

  The static and the non-static solutions correspond to $\epsilon =1$ and $%
\epsilon =-1$, respectively. We restrict our attention  the non-static
case as the results for the static case can be obtained consequently. For $%
\epsilon =-1$, the above equations take the forms
\begin{eqnarray}
ds^{2} &=&dt^{2}-A^{2}(t)dx^{2}-B^{2}(t)dy^{2}-B^{2}(t)\Sigma
^{2}(y,k)dz^{2}, (Metric-I) \\
ds^{2} &=&dt^{2}-A^{2}(t)dx^{2}-B^{2}(t)e^{2x}dy^{2}-B^{2}(t)e^{2x}dz^{2}, (Metric-II)\\
ds^{2} &=&dt^{2}-A^{2}(t)dx^{2}-B^{2}(t)dy^{2}-\{A^{2}(t)\Lambda ^{2}(y,k)
\nonumber \\
&+&B^{2}(t)\Sigma ^{2}(y,k)\}dz^{2}+2A^{2}(t)\Lambda (y,k)dxdz.~~~(Metric-III).
\end{eqnarray}%
The metrics in Eq. (7) become Bianchi type $I(BI)$ or $VII_{0}$ $(BVII_{0})$ for $%
k=0$, $III$ $(BIII)$ for $k=-1$, and Kantowski-Sachs (KS) for $k=+1$. The
metrics in Eq. (8) represent Bianchi type $V(BV)$ or $VII_{h}$ $(BVII_{h})$ metrics.
The metrics in Eq. (9) turn out to be Bianchi types $II(BII)$ for $k=0$, $%
VIII(BVIII)$ or $III(BIII)$ for $k=-1$, and $IX(BIX)$ for $k=+1$. Now, we
will discuss the energy-momentum distribution for the three possible cases
arising from the metric in Eq. (19) for different values of $k$ and from the metric in Eq. (20).
\subsection{Energy-Momentum Densities of the Metric-I}

\quad In this section, we explore the energy-momentum distribution of the metric
in Eq. (19) by using the Einstein, Landau-Lifshitz and Bergmann-Thomson and M$\ddot{o}$%
ller Prescriptions. Three cases are $k=0,k=1,$ and $k=-1$\newline
\textbf{CaseI} ( $k=0$): In this case, the tetrad of the metric in Eq. (19)
can be written as
\begin{equation}
{h^{a}}_{\mu }=\left[
\begin{array}{cccc}
1 & 0 & 0 & 0 \\
0 & A(t) & 0 & 0 \\
0 & 0 & B(t) & 0 \\
0 & 0 & 0 & yB(t)
\end{array}%
\right],
\end{equation}%
and its inverse becomes
\begin{equation}
{h_{a}}^{\mu }=\left[
\begin{array}{cccc}
1 & 0 & 0 & 0 \\
0 & \frac{1}{A(t)} & 0 & 0 \\
0 & 0 & \frac{1}{B(t)} & 0 \\
0 & 0 & 0 & \frac{1}{yB(t)}
\end{array}%
\right] .
\end{equation}%
Here,
\[
h=\det h_{\mu }^{a}=\sqrt{-g}=A(t)B^{2}(t)y.
\]
Using Eqs. (22) and (23) in Eq. (2), we get the following non-zero
components of the Weitzenb$\ddot{o}$ck connections:
\begin{equation}
{\Gamma ^{1}}_{10}=\frac{A^{\prime }(t)}{A(t)},\quad {\Gamma ^{2}}_{20}=%
\frac{B^{\prime }(t)}{B(t)},\quad {\Gamma ^{3}}_{30}=\frac{B^{\prime }(t)}{%
B(t)},\quad {\Gamma ^{3}}_{32}=\frac{1}{y}.
\end{equation}%
The corresponding non-vanishing components of the torsion tensor
found by using Eq. (24) in Eq. (4) in contravariant form are
\begin{eqnarray}
T^{110} &=&\frac{A^{\prime }(t)}{A^{3}(t)}=-T^{101},\quad T^{220}=\frac{%
B^{\prime }(t)}{B^{3}(t)}=-T^{202},  \nonumber \\
T^{330} &=&\frac{B^{\prime }(t)}{y^{2}B^{3}(t)}=-T^{303},\quad T^{332}=-%
\frac{1}{y^{3}B^{4}(t)}=-T^{323}.
\end{eqnarray}
Making use of Eq. (25) in Eq. (9) and then multiplying by $g_{\mu
\nu }$, we find the nonzero components of the $S$ tensor, in mixed form, to be
\begin{eqnarray}
{S_{0}}^{02} &=&-\frac{1}{2yB^{2}(t)}={S_{1}}^{12},\quad
{S_{1}}^{01} =-\frac{B^{\prime }(t)}{B(t)},  \nonumber \\
{S_{2}}^{02} &=&-\frac{(A^{\prime }(t)B(t)+A(t)B^{\prime }(t))}{2A(t)B(t)}={%
S_{3}}^{03}.
\end{eqnarray}%
Substituting Eq. (26) in Eq. (8) yields the required non-vanishing components
of Freud's superpotential as
\begin{eqnarray}
{U_{0}}^{02} &=&-\frac{A(t)}{2}={U_{1}}^{12},\quad
{U_{1}}^{01} ={-yA(t)B^{\prime }(t)B(t)},  \nonumber \\
{U_{2}}^{02} &=&-\frac{yB(t)(B(t)A^{\prime }(t)+A(t)B^{\prime }(t)}{2}={U_{3}%
}^{03}.
\end{eqnarray}%

 If the values from Eq. (27) are substituted in Eq. (7), the non-vanishing
energy and momentum density components can be found  in TPT by using the Einstein,
Landau-Lifshitz and Bergmann-Thomson prescriptions. For the \textbf{\textit{Einstein}} \textit{prescription}, the
non-zero component of the momentum turns out to be
\begin{eqnarray}
hE_{2}^{0} &=&\frac{B(t)A^{\prime }(t)+A(t)B^{\prime }(t)}{8\pi B(t)},
\end{eqnarray}
For the \textbf{\textit{Landau-Lifshitz}} \ \textit{prescription}, the
existing components of the energy and the momentum are
\begin{eqnarray}
hL^{00} &=&-\frac{A^{2}(t)B^{2}(t)}{8\pi },  \nonumber \\
hL^{20} &=&\frac{yA(t)B(t)(A^{\prime }(t)B(t)+A(t)B^{\prime }(t))}{4\pi }.
\end{eqnarray}
For the \textbf{\textit{Bergmann-Thomson}} \textit{prescription}, the surviving momentum component is
\begin{eqnarray}
hB^{20} &=&\frac{A^{\prime }(t)B(t)+A(t)B^{\prime }(t)}{8\pi B(t)}.
\end{eqnarray}

Now, we explore the energy-momentum distribution by using the TP version of the
\textbf{M$\ddot{o}$ller } \textit{Prescription}. For this purpose, we evaluate the
non-vanishing components of the contorsion tensor by using Eq. (25)
in Eq. (6) in contravariant:
\begin{eqnarray}
K^{101}&=& \frac{A^\prime(t)}{A^3(t)} =-K^{011},\quad K^{202}= \frac{%
B^\prime(t)}{B(t)} =-K^{022},  \nonumber \\
K^{303}&=& \frac{B^\prime(t)}{B(t)}=-K^{033},\quad K^{323}= - \frac{1}{y^3
B^4 (t)} =-K^{233}.
\end{eqnarray}
Making use of the Eq. (25) in Eq. (14), we have the non-zero components of the
basic vector part as
\begin{eqnarray}
{V^0}=-\frac{A^\prime(t)B(t)+2A(t)B^\prime(t)}{A(t)B(t)},~~~~~~~~~~ {V^2}=%
\frac{1}{B^2(t)y}.
\end{eqnarray}
The required non-vanishing components of the superpotential in M$\ddot{o}$
ller's tetrad theory can be easily evaluated from Eq. (11) as
\begin{eqnarray}
{U_1}^{01}&=&-\frac{2A(t)B^\prime(t)B(t)y}{\kappa},\quad
{U_0}^{02}=-\frac{A(t)}{\kappa},\quad
{U_1}^{21}=\frac{A(t)}{\kappa}={U_0}^{20},  \nonumber \\
{U_2}^{02}&=&-\frac{B(t)y(A^\prime(t)B(t)+A(t)B^\prime(t))}{\kappa}={U_3}%
^{03},  \nonumber \\
{U_3}^{23}&=&-\frac{\lambda A(t)}{\kappa}.
\end{eqnarray}
If we substitute the values from Eq. (33) in Eq. (15) and then take $c,G=1$
(gravitational units), the energy and momentum density components turn out
to be
\begin{eqnarray}
\Xi_2^0&=&\frac{-B(t)(A^\prime(t)B(t)+A(t)B^\prime(t)}{\kappa}.
\end{eqnarray}%
The above results are summarized in the  Table 1.
\newpage
\textbf{\scriptsize Table 1.}
 {\scriptsize Energy-momentum density
components for different prescriptions}
\begin{center}
\begin{tabular}{|l|l|l|}
\hline \textbf{P} & \textbf{Energy Density} & \textbf{Momentum Density}
\\ \hline ES & $~~hE^{00}=0~~$ &
$hE^{20}=\frac{A^{\prime }(t)B(t)+A(t)B^{\prime }(t)}{8\pi B(t)}$
\\ \hline LL & $hL^{00}=-\frac{A^{2}(t)B^{2}(t)}{8\pi }~$ &
$hL^{20}= \frac{A(t)B(t)[A^{\prime }(t)B(t)+A(t)B^{\prime
}(t)]y}{4\pi }$
\\ \hline BT & $hB^{00}=0$ &
$hB^{20}=\frac{A(t)B^{\prime }(t)+A^{\prime }(t)B(t)}{8\pi B(t)}$
\\ \hline MR & $\Xi ^{00}=0$ & $\Xi
^{20}=\frac{A^{\prime }(t)B(t)+A(t)B^{\prime }(t)}{8\pi B(t)}$
\\ \hline
\end{tabular}
\end{center}
\vspace{0.25cm}

\textbf{CaseII} ($k=1$): In this case, we follow the procedure of
caseI and obtain the energy-momentum distribution for the four
prescriptions, namely, the Einstein, Landau-Lifshitz and Bergmann-Thomson and M$%
\ddot{o}$ller prescriptions. For the \textbf{\textit{Einstein}} \textit{prescription}, the components of the
energy and the momentum are
\begin{eqnarray}
hE_{0}^{0} &=&\frac{A(t)siny}{8\pi },\quad
hE_{2}^{0} =-\frac{B(t)cosy(B(t)A^{\prime }(t)+A(t)B^{\prime }(t))}{8\pi }.
\end{eqnarray}
For the \textbf{\textit{Landau-Lifshitz}} \textit{prescription}, the components
of the energy and the momentum are
\begin{eqnarray}
hL^{00} &=&-\frac{A^{2}(t)B^{2}(t)cos2y}{8\pi },  \nonumber \\
hL^{20} &=&\frac{A(t)B(t)(A^{\prime }(t)B(t)+A(t)B^{\prime }(t))sin2y}{8\pi }.
\end{eqnarray}
For the \textbf{\textit{Bergmann-Thomson}} \textit{prescription}, the components
of the energy and the momentum are
\begin{eqnarray}
hB^{00} &=&\frac{A(t)siny}{8\pi },  \nonumber \\
hB^{20} &=&\frac{cosy(B(t)A^{\prime }(t)+A(t)B^{\prime }(t))}{8\pi B(t)}.
\end{eqnarray}
For the \textbf{\textit{M$\ddot{o}$ller}} \textit{prescription,} the components
of the energy and the momentum are
\begin{eqnarray}
\Xi _{0}^{0} &=&\frac{A(t)siny}{\kappa },  \nonumber \\
\Xi _{2}^{0} &=&-\frac{B(t)cosy(B(t)A^{\prime }(t)+A(t)B^{\prime }(t))}{2}.
\end{eqnarray}
The above results are summarized in the Table 2.
\newpage
\textbf{\scriptsize Table 2.} {\scriptsize Energy-momentum density
components for different prescriptions}
\begin{center}
\begin{tabular}{|l|l|l|}
\hline \textbf{P} & \textbf{Energy Density} & \textbf{Momentum Density}
\\ \hline ES & $hE^{00}=\frac{A(t)\texttt{siny}}{8\pi }$ &
$hE^{20}=\frac{
\texttt{cosy}[B(t)A^{\prime }(t)+A(t)B^{\prime }(t)]}{8\pi B(t)}$ \\
\hline LL & $hL^{00}=-\frac{A^{2}(t)B^{2}(t)\texttt{cos2y}}{8\pi
}$ & $hL^{20}=
\frac{A(t)B(t)[A^{\prime }(t)B(t)+A(t)B^{\prime }(t)]\texttt{sin2y}}{8\pi }$ \\
\hline BT & $hB^{00}=\frac{A(t)\texttt{siny}}{8\pi }$ &
$hB^{20}=\frac{ \texttt{cosy}[B(t)A^{\prime }(t)+A(t)B^{\prime
}(t)]}{8\pi B(t)}$
\\ \hline MR & $\Xi ^{00}=\frac{A(t)\texttt{siny}}{8\pi }$ & $\Xi
^{20}=\frac{\texttt{cosy}[A^\prime (t)B(t)+A(t)B^\prime
(t)]}{8\pi B(t)}$
\\ \hline
\end{tabular}
\end{center}
\vspace{0.25cm}

\textbf{CaseIII }($k=-1$): In this case, we follow the procedure of
caseI and obtain the energy-momentum distribution for the four
prescriptions, namely, the Einstein, Landau-Lifshitz and Bergmann-Thomson and M$%
\ddot{o}$ller prestiptions. For the  \textbf{\textit{Einstein}} \textit{prescription}, the
components of the energy and the momentum are
\begin{eqnarray}
hE_{0}^{0} &=&-\frac{A(t)sinhy}{8\pi },  \nonumber \\
hE_{2}^{0} &=&-\frac{B(t)coshy(B(t)A^{\prime }(t)+A(t)B^{\prime }(t))}{8\pi }.
\end{eqnarray}
For the \textbf{\textit{Landau-Lifshitz}} \ \textit{prescription}, the
components of the energy and the momentum are
\begin{eqnarray}
hL^{00} &=&=\frac{A^{2}(t)B^{2}(t)(cosh^{2}y-sinh^{2}y)}{8\pi },  \nonumber
\\
hL^{20} &=&\frac{A(t)B(t)(A^{\prime }(t)B(t)+A(t)B^{\prime }(t))sinh2y}{8\pi
}.
\end{eqnarray}
For the \textbf{\textit{Bergmann-Thomson}} \textit{prescription}, the components
of the energy and the momentum are
\begin{eqnarray}
hB^{00} &=&-\frac{A(t)sinhy}{8\pi },  \nonumber \\
hB^{20} &=&\frac{(A^{\prime }(t)B(t)+A(t)B^{\prime }(t))coshy}{8\pi B(t)}.
\end{eqnarray}
For the \textbf{\textit{M$\ddot{o}$ller}} \textit{prescription}, the components
of the energy and the momentum are
\begin{eqnarray}
\Xi _{0}^{0} &=&-\frac{B(t)sinhy}{\kappa },  \nonumber \\
\Xi _{2}^{0} &=&-\frac{B(t)coshy(B(t)A^{\prime }(t)+A(t)B^{\prime }(t))}{%
\kappa }.
\end{eqnarray}
The above results are summarized in the  Table 3.
\newpage
\textbf{\scriptsize Table 3.} {\scriptsize Energy-momentum density
components for different prescriptions}.
\begin{center}
\begin{tabular}{|l|l|l|}
\hline \textbf{P} & \textbf{Energy Density} & \textbf{Momentum Density}
\\ \hline ES & $hE^{00}=-\frac{A(t)\texttt{sinhy}}{8\pi }$ &
$hE^{20}=\frac{
\texttt{coshy}[B(t)A^{\prime }(t)+A(t)B^{\prime }(t)]}{8\pi B(t) }$ \\
\hline LL & $hL^{00}=-\frac{A^{2}(t)B^{2}(t)\texttt{cosh2y}}{8\pi
}$ & $hL^{20}=\frac{A(t)B(t)[A^{\prime }(t)B(t)+A(t)B^{\prime
}(t)]\texttt{sinh2y}}{ 8\pi }$ \\ \hline BT &
$hB^{00}=-\frac{A(t)\texttt{sinhy}}{8\pi }$ & $hB^{20}=\frac{
[A^{\prime }(t)B(t)+A(t)B^{\prime }(t)]\texttt{coshy}}{8\pi B(t)}$
\\ \hline MR & $\Xi ^{00}=-\frac{A(t)\texttt{sinhy}}{8\pi }$ & $\Xi
^{20}=\frac{\texttt{coshy}[A^\prime (t)B(t)+A(t)B^\prime
(t)]}{8\pi B(t) }$
\\ \hline
\end{tabular}
\end{center}
\subsection{Energy-Momentum Densities of the Metric-II}

\quad The tetrad of the metric in Eq. (20) can be written as
\begin{equation}
{h^{a}}_{\mu }=\left[
\begin{array}{cccc}
1 & 0 & 0 & 0 \\
0 & A(t) & 0 & 0 \\
0 & 0 & e^{x}B(t) & 0 \\
0 & 0 & 0 & e^{x}B(t)
\end{array}%
\right],
\end{equation}%
and its inverse becomes
\begin{equation}
{h_{a}}^{\mu }=\left[
\begin{array}{cccc}
1 & 0 & 0 & 0 \\
0 & \frac{1}{A(t)} & 0 & 0 \\
0 & 0 & \frac{1}{e^{x}B(t)} & 0 \\
0 & 0 & 0 & \frac{1}{e^{x}B(t)}%
\end{array}%
\right] .
\end{equation}%
Here
\[
h=\det h_{\mu }^{a}=\sqrt{-g}=A(t)B^{2}(t)e^{2x}.
\]
Using Eqs. (43) and (44) in Eq. (2), we get the following non-zero
components of the Weitzenb$\ddot{o}$ck connections
\begin{eqnarray}
{\Gamma ^{1}}_{10} &=&\frac{A^{\prime }(t)}{A(t)},\quad {\Gamma ^{2}}_{20}=%
\frac{B^{\prime }(t)}{B(t)},\quad {\Gamma ^{3}}_{30}=\frac{B^{\prime }(t)}{%
B(t)},  \nonumber \\
{\Gamma ^{3}}_{31} &=&{1},\quad {\Gamma ^{2}}_{21}={1}.
\end{eqnarray}%
The corresponding non-vanishing components of the torsion tensor in contravariant form are
\begin{eqnarray}
T^{110} &=&\frac{A^{\prime }(t)}{A^{3}(t)}=-T^{101},\quad T^{220}=\frac{%
B^{\prime }(t)}{e^{2x}B^{3}(t)}=-T^{202},  \nonumber \\
T^{330} &=&\frac{B^{\prime }(t)}{e^{2x}B^{3}(t)}=-T^{303},\quad T^{331}=-%
\frac{-1}{e^{2x}A^{2}(t)B^{2}(t)}=-T^{313},  \nonumber \\
T^{221} &=&-\frac{-1}{e^{2x}A^{2}(t)B^{2}(t)}=-T^{212}.
\end{eqnarray}
Making use of Eq. (46) in Eq. (9) and then multiplying by $g_{\mu
\nu }$, we find the non-zero components of the $S$ tensor, in mixed form, to be
\begin{eqnarray}
{S_{0}}^{01} &=&-\frac{1}{A^{2}(t)},\quad
{S_{1}}^{00} =-\frac{A(t)(A^{\prime }(t)B(t)+A(t)B^{\prime }(t))}{2B(t)},
\nonumber \\
{S_{1}}^{01} &=&\frac{(A(t)(A^{\prime }(t)+2)}{2A^{4}(t)}.
\end{eqnarray}%
Substituting Eq. (47) in Eq. (8) yields the required non-vanishing components
of  Freud's superpotential:
\begin{eqnarray}
{U_{0}}^{01} &=&-\frac{e^{2x}B^{2}(t)}{A(t)},\quad
{U_{1}}^{01} =-{e}^{2x}{A(t)B^{\prime }(t)B(t)},\nonumber \\
{U_{2}}^{02} &=&-\frac{{e}^{2x}B(t)(B(t)A^{\prime }(t)+A(t)B^{\prime }(t)}{2}%
={U_{3}}^{03}  \nonumber \\
{U_{2}}^{12} &=&\frac{e^{2x}B^{2}(t)}{2A(t)}={U_{3}}^{13}.
\end{eqnarray}%
When the values from Eq. (48) are substituted in Eq. (7), the required
non-vanishing energy and momentum density components in TPT can be found by using the  Einstein's,
Landau-Lifshitz and Bergmann-Thomson prescriptions, respectively, as
\begin{eqnarray}
hE_{0}^{0} &=&-\frac{e^{2x}B^{2}(t)}{2\pi A(t)},\quad
hE_{1}^{0} =\frac{e^{2x}B(t)B^{\prime }(t)}{2\pi A(t)},
\\
hL^{00} &=&-\frac{e^{4x}B^{4}(t)}{\pi },\quad
hL^{10} =\frac{e^{4x}B^{3}(t)B^{\prime }(t)}{\pi }
\end{eqnarray}%
and
\begin{eqnarray}
hB^{00} &=&-\frac{e^{2x}B^{2}(t)}{2\pi A(t)},\quad
hB^{10} =\frac{e^{2x}B(t)B^{\prime }(t)}{2\pi A(t)}.
\end{eqnarray}

  Now, we explore the energy-momentum distribution by using the TP version of the
\textbf{M$\ddot{o}$ller \textit{Prescription}}. For this purpose, we evaluate the
non-vanishing components of the contorsion tensor by using Eq. (46) in Eq. (6)
in contravarient form as
\begin{eqnarray}
K^{101} &=&\frac{A^{\prime }(t)}{A^{3}(t)}=-K^{011},\quad K^{202}=\frac{%
B^{\prime }(t)}{e^{2x}B(t)}=-K^{022},  \nonumber \\
K^{303} &=&\frac{B^{\prime }(t)}{e^{2x}B^{3}(t)}=-K^{033},\quad K^{313}=-%
\frac{1}{e^{2x}A^{2}(t)B^{2}(t)}=-K^{133},  \nonumber \\
K^{122} &=&\frac{1}{e^{2x}A^{2}(t)B^{2}(t)}=-K^{212}.
\end{eqnarray}%
The non-vanishing components of the basic vectors are evaluated, and the vector
part is
\[
{V^{0}}=-\frac{A^{\prime }(t)B(t)+2A(t)B^{\prime }(t)}{A(t)B(t)},~~~~~~~~~~{%
V^{1}}=\frac{2}{A^{2}(t)}.
\]%
The required non-vanishing components of the superpotential, in M$\ddot{o}$%
ller's tetrad theory, are
\begin{eqnarray}
{U_{0}}^{01} &=&\frac{-2e^{2x}B^{2}(t)}{A(t)\kappa },\quad
{U_{1}}^{01} =-\frac{2A(t)B^{\prime }(t)B(t)e^{2x}}{\kappa },  \nonumber \\
{U_{0}}^{02} &=&-\frac{A(t)}{\kappa },\quad
{U_{2}}^{02} =-\frac{e^{2x}B(t)(A^{\prime }(t)B(t)+A(t)B^{\prime }(t))}{%
\kappa }={U_{3}}^{03},  \nonumber \\
{U_{3}}^{13} &=&\frac{3e^{2x}B^{2}(t)}{A(t)\kappa }={U_{2}}^{12}.
\end{eqnarray}
Substituting these results in Eq. (15) and using $c,G=1$ it yield the energy and
the momentum densities in M$\ddot{o}$ller's prescription:
\begin{eqnarray}
\Xi _{0}^{0} &=&\frac{-4B^{2}(t)e^{2x}}{A(t)\kappa },\quad
\Xi _{1}^{0} =\frac{-4A(t)B(t)B^{\prime }(t)e^{2x}}{\kappa }.
\end{eqnarray}
The above results are summarized in the Table 4.

\textbf{\scriptsize Table 4.} {\scriptsize Energy-momentum densities
components for Different Prescriptions}.
\begin{center}
\begin{tabular}{|l|l|l|}
\hline \textbf{P} & \textbf{Energy Density} & \textbf{Momentum Density}
\\ \hline ES & $~~hE^{00}=-\frac{e^{2x}B^{2}(t)}{2\pi
A(t)}$ & $~hE^{10}= \frac{e^{2x}B(t)B^{\prime }(t)}{2\pi A(t)}$ \\
\hline LL & $hL^{00}=-\frac{e^{4x}B^{4}(t)}{\pi
}~~~$ & $~hL^{10}= \frac{e^{4x}B^{3}(t)B^{\prime }(t)}{\pi }$ \\
\hline BT & $hB^{00}=-\frac{e^{2x}B^{2}(t)}{2\pi A(t)}$ &
$hB^{10}= \frac{e^{2x}B(t)B^{\prime }(t)}{2\pi A(t)}$
\\ \hline MR & $\Xi
^{00}=-\frac{e^{2x}B^{2}(t)}{A(t)2\pi }$ & $\Xi
^{10}=\frac{e^{2x}B(t)B^{\prime }(t)}{2\pi A(t)}$ \\
\hline
\end{tabular}
\end{center}

\subsection{Energy-Momentum Densities of the Metric-III}

\quad In this section, we explore the energy-momentum distribution of
the metric in Eq. (21) by using the TP version of the Einstein,
Landau-Lifshitz and Bergmann-Thomson prescriptions for the
three cases: $ \alpha).~ k=0, \beta),~ k=1, $ and $ \gamma ),~k=-1$.\\
\textbf{Case}$\alpha$: ($k=0$): In this case, the tetrad of
the metric in Eq. (21) can be written as
\begin{equation}
{h^{a}}_{\mu }=\left[
\begin{array}{cccc}
1 & 0 & 0 & 0 \\
0 & A(t) & 0 & \frac{y^2A(t)}{2} \\
0 & 0 & B(t) & 0 \\
0 & 0 & 0 & yB(t)
\end{array}
\right],
\end{equation}
and its inverse turns out to be
\begin{equation}
{h_{a}}^{\mu }=\left[
\begin{array}{cccc}
1 & 0 & 0 & 0 \\
0 & \frac{1}{A(t)} & 0 & 0 \\
0 & 0 & \frac{1}{B(t)} & 0 \\
0 & \frac{-y}{2B(t)} & 0 & \frac{1}{yB(t)}
\end{array}
\right].
\end{equation}
Here,
$$
h=\det{h^{a}}_{\mu}=\sqrt{-g}=A(t)B^{2}(t)y.
$$

  Using Eqs. (55) and (56) in Eq. (2), we get the
following non-zero components of the Weitzenb$\ddot{o}$ck
connections:
\begin{eqnarray}
{\Gamma^{1}}_{10}&=&\frac{A^{\prime}(t)}{A(t)},\quad
{\Gamma^{1}}_{30}=\frac{y^2(A^{\prime}(t)B(t)-A(t)B^{\prime }(t))}{2A(t)B(t)}, \nonumber \\
{\Gamma^{2}}_{20}&=&\frac{B^{\prime}(t)}{B(t)},\quad
{\Gamma^{3}}_{30}=\frac{B^{\prime}(t)}{B(t)},\quad
{\Gamma^{3}}_{32}=\frac{1}{y}.
\end{eqnarray}
The corresponding non-vanishing components of the torsion tensor
can be found, by using Eq. (57) in Eq. (4), which is
antisymmetric in nature in contravariant form  as
\begin{eqnarray}
{T^{110}}&=&\frac{4B^3(t)A^{\prime }(t)+y^2A^3(t)B^{\prime }(t)}{4A^3(t)B^3(t)}=-T^{101},  \nonumber \\
{T^{103}}&=&\frac{B^{\prime }(t)}{2B^{3}(t)}=-T^{130},\quad
{T^{220}}=\frac{B^{\prime }(t)}{B^{3}(t)}=-T^{202},  \nonumber \\
{T^{330}}&=&\frac{B^{\prime }(t)}{y^{2}B^{3}(t)}=-T^{303},\quad
{T^{112}}=\frac{y}{4B^{4}(t)}=-T^{121},  \nonumber \\
{T^{312}}&=&\frac{1}{2yB^{4}(t)}=-T^{321},\quad
{T^{123}}=\frac{1}{2yB^{4}(t)}=-T^{132},  \nonumber \\
{T^{301}}&=&\frac{B^{\prime }(t)}{2B^{3}(t)}=-T^{301},\quad
{T^{332}}=-\frac{1}{y^{2}B^{4}(t)}=T^{323}.
\end{eqnarray}

   Using the same procedure as used for the  metric in Eq. (19), we have evaluated
the non-vanishing energy-momentum density components for the
Einstein Landau-Lifshitz and Bergmann-Thomson prescription (in
TPT) as
\begin{eqnarray}
hE_{2}^{0}&=&-\frac{B(t)(B(t)A^{\prime }(t)+A(t)B^{\prime }(t))}{8\pi },
\\
hL^{00}&=&-\frac{A^{2}(t)B^{2}(t)}{8\pi },  \nonumber \\
hL^{20}&=&\frac{y A(t)B(t)(A^{\prime }(t)B(t)+A(t)B^{\prime}(t))}{4\pi }
\end{eqnarray}
and
\begin{eqnarray}
hB^{20}&=&\frac{A^{\prime
}(t)B(t)+A(t)B^{\prime }(t)}{8\pi B(t)}.
\end{eqnarray}
Now, we explore the energy-momentum distribution by using the TP version of the
\textbf{M$\ddot{o}$ller} \textit{prescription}. For this purpose, we evaluate the
non-vanishing components of the contorsion tensor by using Eq. (58)
in Eq. (6) in contravariant form as
\begin{eqnarray}
K^{101}&=& \frac{A^\prime(t)}{A^3(t)}+\frac{y^{2}B'(t)}{4B^{3}(t)} =-K^{011},\quad
K^{202}= \frac{B^\prime(t)}{B^{3}(t)} =-K^{022},  \nonumber \\
 K^{301}&=& -\frac{%
B^\prime(t)}{{2}B^{3}(t)} =-K^{103},\quad
K^{303}= \frac{B^\prime(t)}{y^{2}B^{3}(t)}=-K^{033}, \nonumber \\
K^{323}&=& - \frac{1}{y^2B^4 (t)} =-K^{233},\quad
K^{211}=\frac{-y}{4B^{4}(t)}=-K^{121}, \nonumber \\
K^{312}&=&-\frac{1}{2yB^{4}(t)}=-K^{132}.
\end{eqnarray}
Making use of Eq. (58) in Eq. (14), we have the non-zero components of the
basic vector part:
\begin{eqnarray}
{V^0}=-\frac{A^\prime(t)B(t)+2A(t)B^\prime(t)}{A(t)B(t)},~~~~~~~~~~ {V^2}=%
\frac{1}{B^2(t)y}.
\end{eqnarray}

  The required non-vanishing components of the superpotential in M$\ddot{o}$%
ller's tetrad theory can be easily evaluated from Eq. (11) as
\begin{eqnarray}
{U_1}^{01}&=&-\frac{2A(t)B^\prime(t)B(t)y}{\kappa}=-{U_1}^{10},\quad
{U_0}^{02}=-\frac{A(t)}{\kappa}=-U_{0}^{20},  \nonumber \\
{U_3}^{01}&=&y^{2}B(t)(\frac{B(t)(A^\prime(t)-A(t)B^\prime(t))}{2\kappa})=-{U_3}%
^{10},  \nonumber \\
{U_2}^{02}&=&-\frac{B(t)y(A^\prime(t)B(t)+A(t)B^\prime(t))}{\kappa}={U_3}%
^{03},  \nonumber \\
{U_3}^{23}&=&-\frac{\lambda A(t)}{\kappa},\quad
{U_1}^{21}=\frac{A(t)}{\kappa}={U_0}^{20}, \nonumber \\
{U_3}^{12}&=&-y^{2}A(t)(\frac{y^{2}(1+\lambda)(A^{2}(t)+4\lambda B^{2}(t)}{8\kappa B^{2}(t)})=-{U_3}%
^{21},  \nonumber \\
{U_1}^{21}&=&A(t)(\frac{y^{2}(1+\lambda)(A^{2}(t)+4 B^{2}(t)}{4\kappa B^{2}(t)})=-{U_1}%
^{12},  \nonumber \\
{U_2}^{13}&=&\frac{(-1+\lambda)A(t)}{2\kappa}=-{U_2}^{31},\quad
{U_1}^{23}=-\frac{(1+\lambda)A^{2}(t)}{2\kappa B^{2}(t)}=-{U_1}^{32}, \nonumber \\
{U_3}^{23}&=&-\frac{y^{2}(1+\lambda)A^{2}(t)}{4\kappa B^{2}(t)}=-{U_3}^{32}.
\end{eqnarray}
Substituting the values from Eq. (64) in Eq. (15) and then taking $c,G=1$
(gravitational units), we find the energy and momentum density components
to be
\begin{eqnarray}
\Xi_2^0&=&\frac{-B(t)(A^\prime(t)B(t)+A(t)B^\prime(t)}{\kappa}.
\end{eqnarray}%
\newline
The above results are summarized in the Table 5.\\\\\\\\\\\\

\textbf{\scriptsize Table 5.} {\scriptsize Energy-momentum density
components in different prescriptions}.
\begin{center}
\begin{tabular}{|l|l|l|}
\hline \textbf{P} & \textbf{Energy Density} & \textbf{Momentum Density}
\\ \hline ES & $~~hE^{00}=0~~$ &
$hE^{20}=\frac{A^{\prime }(t)B(t)+A(t)B^{\prime }(t)}{8\pi B(t)}$
\\ \hline LL & $hL^{00}=-\frac{A^{2}(t)B^{2}(t)}{8\pi }~$ &
$hL^{20}= \frac{A(t)B(t)[A^{\prime }(t)B(t)+A(t)B^{\prime
}(t)]y}{4\pi }$
\\ \hline BT & $hB^{00}=0$ &
$hB^{20}=\frac{A(t)B^{\prime }(t)+A^{\prime }(t)B(t)}{8\pi B(t)}$
\\ \hline MR & $\Xi ^{00}=0$ & $\Xi
^{20}=\frac{A^\prime (t)B(t)+A(t)B^\prime (t)}{8\pi B(t) }$
\\ \hline
\end{tabular}
\end{center}
\vspace{0.25cm}

\textbf{Case}$\beta$: ($k=1$): In this case, we follow the same procedure as was used for
caseI and obtain the energy-momentum distribution for the four
prescriptions, namely, the Einstein, Landau-Lifshitz and Bergmann-Thomson prescriptions.
For the  \textbf{\textit{Einstein }} \ \textit{prescription}, the
components of the energy and the momentum turn out to be
\begin{eqnarray}
hE_{2}^{0}&=&-\frac{B(t)(B(t)A^{\prime }(t)+A(t)B^{\prime }(t))cosy}{8\pi },\nonumber \\
hE_{0}^{0}&=&\frac{A(t)siny}{8\pi }.
\end{eqnarray}
For the  \textbf{\textit{Landau-Lifshitz}} \ \textit{prescription}, the
components of the energy and the momentum are
\begin{eqnarray}
hL^{00}&=&0,\\
hL^{20}&=&\frac{A(t)B(t)\texttt{siny}(A^{\prime}(t)B(t)+A(t)B^{\prime}(t))cosy}{4\pi }.
\end{eqnarray}
For the \textbf{\textit{Bergmann-Thomson}} \textit{prescription}, the components
of the energy and the momentum are given as
\begin{eqnarray}
hB^{20}&=&\frac{(A^{\prime }(t)B(t)+A(t)B^{\prime }(t))cosy}{8\pi B(t)}, \nonumber \\
hB^{00}&=&\frac{A(t)siny}{8\pi}.
\end{eqnarray}
For the  \textbf{\textit{M$\ddot{o}$ller}} \textit{prescription}, the non vanishing components
of the energy and the momentum are
\begin{eqnarray}
\Xi_2^0&=&-\frac{B(t)(B(t)A^{\prime }(t)+A(t)B^{\prime }(t))cosy}{\kappa}.
\end{eqnarray}%
The above results are summarized in the Table 6.

\textbf{\scriptsize Table 6.} {\scriptsize Energy-momentum density
components in different prescriptions}
\begin{center}
\begin{tabular}{|c|c|c|}
\hline \textbf{P} & \textbf{Energy Density} & \textbf{Momentum Density}\\ \hline
ES      & $hE^{00}=\frac{A(t)siny}{8\pi }$ & $hE^{20}=\frac{cosy[B(t)A'(t)+A(t)B'(t)]}{8\pi B(t)}$ \\ \hline
LL     & $hL^{00}=0$ & $hL^{20}=\frac{1}{4\pi }\{A(t)B(t)siny$\\&&$~~~~~~~~~(A'(t)B(t)+A(t)B'(t))cosy\}$ \\ \hline
BT      & $hB^{00}=\frac{A(t)siny}{8\pi}$ & $hB^{20}=\frac{ cosy[B(t)A'(t)+A(t)B'(t)]}{8\pi B(t)}$ \\ \hline
MR & $\Xi^{00}=0$ &  $\Xi^{20}=\frac{ cosy[B(t)A'(t)+A(t)B'(t)]}{8\pi B(t)}$
\\ \hline
\end{tabular}
\end{center}
\noindent

\textbf{Case}$\gamma$: ($k=-1$): In this case, we follow the same procedure as for
caseI and obtain the energy-momentum distribution for the four
prescriptions, namely, the Einstein, Landau-Lifshitz and Bergmann-Thomson as given prescriptions.
For the \textbf{\textit{Einstein}} \textit{prescription}, the
components of the energy and the momentum are
\begin{eqnarray}
hE_{2}^{0}&=&-\frac{B(t)(B(t)A^{\prime }(t)+A(t)B^{\prime }(t))coshy}{8\pi},\nonumber \\
hE_{0}^{0}&=&-\frac{A(t)sinhy}{8\pi}.
\end{eqnarray}
For the \textbf{\textit{Landau-Lifshitz}} \ \textit{prescription}, the
components of the energy and the momentum are
\begin{eqnarray}
hL^{00}&=&-\frac{A^{2}(t)B^{2}(t) sinh^{2}y}{4\pi },  \nonumber \\
hL^{20}&=&\frac{A(t)B(t)\texttt{sinhy}(A^{\prime}(t)B(t)+A(t)B^{\prime}(t))coshy}{4\pi }.
\end{eqnarray}
For the \textbf{\textit{Bergmann-Thomson}} \textit{prescription}, the components
of the energy and the momentum are
\begin{eqnarray}
hB^{20}&=&\frac{A^{\prime }(t)B(t)+A(t)B^{\prime }(t)coshy}{8\pi B(t)}, \nonumber \\
hB^{00}&=&-\frac{A(t)sinhy}{8\pi}.
\end{eqnarray}
For the \textbf{\textit{M$\ddot{o}$ller}} \textit{prescription}, the non-zero components
of the energy and the momentum are
\begin{eqnarray}
\Xi_2^0&=&-\frac{B(t)(B(t)A^{\prime }(t)+A(t)B^{\prime }(t))coshy}{\kappa}.
\end{eqnarray}%
The above results are summarized in the Table 7.
\newpage
\textbf{\scriptsize Table 7.} {\scriptsize Energy-momentum density
components in different prescriptions}.
\begin{center}
\begin{tabular}{|c|c|c|}
\hline \textbf{P} & \textbf{Energy Density} & \textbf{Momentum Density}
\\ \hline ES & $hE^{00}=-\frac{A(t)sinhy}{8\pi }$ & $hE^{20}=\frac{coshy[B(t)A'(t)+A(t)B'(t)]}{8\pi B(t)}$ \\
\hline LL & $hL^{00}=-\frac{A^{2}(t)B^{2}(t)}{4\pi}sinh^2 y$& $hL^{20}=
\frac{1}{4\pi }\{A(t)B(t)sinhy$\\&
&$~~~~~(A'(t)B(t)+A(t)B'(t))coshy\}$ \\
\hline BT & $hB^{00}=-\frac{A(t)sinhy}{8\pi }$
& $hB^{20}=\frac{ coshy[B(t)A'(t)+A(t)B'(t)]}{8\pi B(t)}$ \\ \hline
MR & $\Xi^{00}=0$ & $\Xi
^{20}=\frac{coshy[B(t)A'(t)+A(t)B'(t)]}{8\pi B(t) }$
\\ \hline
\end{tabular}
\end{center}
\section{SUMMARY AND DISCUSSION}

\quad The problem of localization of energy has been re-considered in the
framework of TPG by many scientists. Some authors [25-28] showed that
energy-momentum can also be localized in this theory. Many examples have been
explored by different researchers, and are available in literature. Two, main conclusions
have been made. Firstly, the results of TPG agree for some prescriptions in some
spacetimes while the same prescriptions yield different results for some other
spacetimes.

Vargas [28] found that the total energy of the closed Friedmann-Robertson-Walker
model was zero by using the TP version of Einstein and Landau-Lifshitz complexes which
agreed with the results of GR [31]. Sharif and his collaborators [24], [33-41] used different
prescriptions to determine the energy-momentum distributions for various
spacetimes and found that the results of TPG and GR were not consistent. Recently,
Amir and his collaborators [42] evaluated the energy-momentum distribution of non-static
plane-symmetric spacetimes by using different prescriptions in the context of TPG and GR
and showed that the results for the Einstein, Landau-Lifshitz, Bergmann-Thomson prescriptions
are the same in both the theories but are different form those obtained when using the M$\ddot{o}$ller
prescription.

Now, we have extended this work for the whole family of LRS
spacetimes. We consider the three metrics representing the LRS
spacetimes and explored the energy-momentum distribution by using
the TP version of the the Einstein, Landau-Lifshitz,
Bergmann-Thomson, and M$\ddot{o}$ller prescriptions.
Three cases arise for metric I and III (three
different values of $k$) while metric II yields only one case.
The energy and the momentum density components of the Einstein,
Landau-Lifshitz, Bergmann-Thomson and the M$\ddot{o}$ller
prescriptions for all seven cases are given in tables 1-7.

We see that energy and the momentum take well-defined and definite forms for
each prescription in all seven cases. Tables $1-7$ show that the momentum density components of
the Einstein, Bergmann-Thomson and M$\ddot{o}$ller prescriptions are the same in all
seven cases of the metrics I, II and III while the Landau-Lifshitz prescription
yields different results. Further, the energy components of all cases of metrics
I and II turn out to be same for the Einstein, Bergmann-Thomson and M$\ddot{o}$ller prescriptions
but the energy density components of both the Landau-Lifshitz and the M$\ddot{o}$ller prescriptions
have been shown to be different for metric-III.

\vspace{1.5cm}
\newpage
{\bf REFERENCES}

\begin{description}
\item {[1]} C.W. Misner, K.S. Thorne  and J.A. Wheeler, \textit{Gravitation%
} (Freeman, New York, 1973).

\item {[2]} A. Trautman, \textit{Gravitation}: \textit{An Introduction to
Current Research}, edetted by L. Witten. (Wiley, New York, 1962).

\item {[3]} L.D. Landau and E.M. Lifshitz,  \textit{The Classical Theory of
Fields} (Addison-Wesley Press, New York, 1962).

\item {[4]} A. Papapetrou, R. Proc. Irish Acad. \textbf{A 52}, 11(1948).

\item {[5]} P.G. Bergman and R. Thomson,  Phys. Rev. \textbf{89}, 400(1958).

\item {[6]} R.C. Tolman,  \textit{Relativity Thermodynamics and Cosmology}
(Oxford University Press, Oxford, 1934).

\item {[7]} S. Weinberg,  \textit{Gravitation and Cosmology} (Wiley, New
York, 1972).

\item {[8]} C. M$\ddot{o}$ller,  Ann. Phys. (N.Y.) \textbf{4}, 347(1958).

\item {[9]} F.I. Cooperstock and R.S. Sarracino,  J. Phys. \textbf{A}: Math. Gen.
\textbf{11}, 877(1978).

\item {[10]} H. Bondi,  Proc. R. Soc. London \textbf{A 427}, 249(1990).

\item {[11]} R. Penrose,  Proc. Roy. Soc. London  \textbf{A 388}, %
457(1982); GR10 Conference eds.B., Bertotti, F. de Felice  and A. Pascolini,
 Padova \textbf{1}, 607(1983).

\item {[12]} J.D. Brown and Jr., York  Phys. Rev. \textbf{D 47},%
1407(1993).

\item {[13]} S.A. Hayward, Phys. Rev. \textbf{D 497}, 831(1994).

\item {[14]} G. Bergqvist, Class. Quantum Gravit. \textbf{9}, 1753(1992).

\item {[15]} C.C. Chang,  J.M. Nester and C. Chen,  Phys. Rev. Lett.
\textbf{83}, 1897(1999).

\item {[16]} S.S. Xulu, Int. J. Mod. Phys. \textbf{A 15}, 2979(2000); Mod.
Phys. Lett. \textbf{A 15}, 1151(2000) and references therein.

\item {[17]} S.S. Xulu,  Astrophys. Space Sci. \textbf{283}, 23(2003).

\item {[18]} K.S. Virbhadra, Phys. Rev. \textbf{D 42}, 2919(1990).

\item {[19]} K.S. Virbhadra, and J.C. Parikh,  Phys. Lett. \textbf{B 317}, %
312(1993).

\item {[20]}  K.S. Virbhadra, and J.C. Parikh,  Phys. Lett. \textbf{B 331}, %
302(1994).

\item {[21]}  N. Rosen, and K.S. Virbhadra,  Gen. Relativ. Gravit. \textbf{25}, %
429(1993).

\item {[22]} J.M. Aguirregabiria, A. Chamorro and K.S. Virbhadra,  Gen.
Relativ. Gravit. \textbf{28}, 1393(1996).
 
\item {[23]} K.S. Virbhadra,  Phys. Rev. \textbf{D 60}, 104041(1999).

\item {[24]} M. Sharif,  Int. J. Mod. Phys. \textbf{A 17}, 1175(2002); \textbf{%
A 18}, 4361(2003); Errata \textbf{A 19}, 1495(2004). Int. J of Mod. Phys. \textbf{%
D 13}, 1019(2004).

\item {[25]} I. Radinschi,  Mod. Phys. Lett. \textbf{A 16}, 673(2001).

\item {[26]} F.I. Mikhail, M.I., Wanas, A. Hindawi,  and E.I. Lashin,  Int.
J. Theor. Phys. \textbf{32}, 1627(1993).

\item {[27]} G.G.L. Nashed,  Phys. Rev. \textbf{D 66}, 060415(2002).

\item {[28]} T. Vargas,  Gen. Relativ. Gravit. \textbf{30}, 1255(2004).

\item {[29]} M. Salti,  A. Havare,  Int. J. of Mod. Phys. \textbf{A 20}, %
2169(2005);  O. Aydogdu, and  M. Salti, Astrophys. Space Sci. \textbf{229}, %
227(2005);  O. Aydogdu, M. Salti and M. Korunur,  Acta Phys. Slov. \textbf{%
55}, 537(2005);  M. Salti, Astrophys. Space Sci. \textbf{229}, 159(2005).

\item {[30]} R.M. Gad,  Astrophys. Space Sci. \textbf{295}, 495(2004).

\item {[31]} N. Rosen, Gen. Relativ. Gravit. \textbf{26}, 319(1994).

\item {[32]} M. Sharif and M.J. Amir, Mod. Phys. Lett. \textbf{A 22}, 425(2007).

\item {[33]} M. Sharif and M.J. Amir, Gen. Relativ. Gravit. \textbf{38}, 1735(2006).

\item {[34]} M. Sharif and M.J. Amir, Gen. Relativ. Gravit.
\textbf{39}, 989(2007).

\item {[35]} M. Sharif and M.J. Amir, Canadian J. Phys. \textbf{86}, 1297(2008).

\item {[36]} M. Sharif and M.J. Amir, Mod. Phys. Lett.
\textbf{A 23}, 3167(2008).

\item {[37]} M. Sharif and M.J. Amir, Canadian J. Phys.
\textbf{86}, 1091(2008).

\item {[38]} M. Sharif and M.J. Amir, Int. J. Theor. Phys. \textbf{47}, 1742(2008).

\item {[39]} M. Sharif, and A. Jawad,  Mod. Phys. Lett. \textbf{A 25}, 3241(2010).

\item{[40]} M. Sharif,  and  S. Taj, Mod. Phys. Lett. \textbf{A 25}, 221(2010).

\item {[41]}  M. Sharif, and S. Taj,  Astrophys. Space Sci. \textbf{325}, 75(2010).
\item {[42]} M.J. Amir,  S. Ali,  and  T. Ismaeel, Chines  J. of  Phys. \textbf{50}, 14(2012).
\end{description}
\end{document}